\documentclass[preprint,aps,floatfix]{revtex4}
\usepackage{booktabs}
\usepackage{graphicx}
\usepackage[ pdftex, plainpages = false, pdfpagelabels,
                 pdfpagelayout = SinglePage,
                 bookmarks,
                 bookmarksopen = true,
                 bookmarksnumbered = true,
                 breaklinks = true,
                 linktocpage=true,
                 pagebackref=false,
                 colorlinks = true,
                 linkcolor = BrickRed,
                 urlcolor  = blue,
                 citecolor = BrickRed,
                 anchorcolor = green,
                 hyperindex = true,
                 hyperfigures
                 ]{hyperref}
\usepackage{multirow}
\usepackage[usenames, dvipsnames]{color}
\makeatletter

\newcommand{\category}[1]{{#1}}

\begin{document}

\title{\href{http://necsi.edu/research/social/nyttwitter/}{An Exploration of Social Identity: The Geography and Politics of News-Sharing Communities in Twitter}}

\author{
    Ama\c{c} Herda\u{g}delen, Wenyun Zuo, Alexander Gard-Murray and \href{http://necsi.edu/faculty/bar-yam.html}{Yaneer Bar-Yam}\\
    \href{http://www.necsi.edu}{New England Complex Systems Institute}\\
    238 Main Street Suite 319, Cambridge, Massachusetts 02142, US
}
\date{\today}

\begin{abstract}
The importance of collective social action in current events is manifest in the Arab Spring and Occupy movements. Electronic social media have become a pervasive channel for social interactions, and a basis of collective social response to information. The study of social media can reveal how individual actions combine to become the collective dynamics of society. Characterizing the  groups that form spontaneously may reveal both how individuals self-identify and how they will act together. 
Here we map  the social, political, and geographical properties of news-sharing communities on Twitter, a popular micro-blogging platform.
We track user-generated messages that contain links to New York Times online articles and we label users according to the topic of the links they share, their geographic location, and their self-descriptive keywords.
When users are clustered based on who follows whom in Twitter, we find social groups separate by whether they are interested in 
local (NY), national (US) or global (cosmopolitan) issues. The national group subdivides into liberal, conservative and other, the latter being a diverse but mostly business oriented group with sports, arts and other splinters. The national political groups are based across the US but are distinct from the national group that is broadly interested in a variety of topics. 
A person who is cosmopolitan associates with others who are cosmopolitan, and a US liberal / conservative associates with others who are US liberal / conservative, creating separated social groups with those identities.  
The existence of ``citizens" of local, national and cosmopolitan communities is a basis for dialog and action at each of these levels of societal organization. 
\end{abstract}

\maketitle
\begin{figure}[b]
\includegraphics[width=0.8\textwidth]{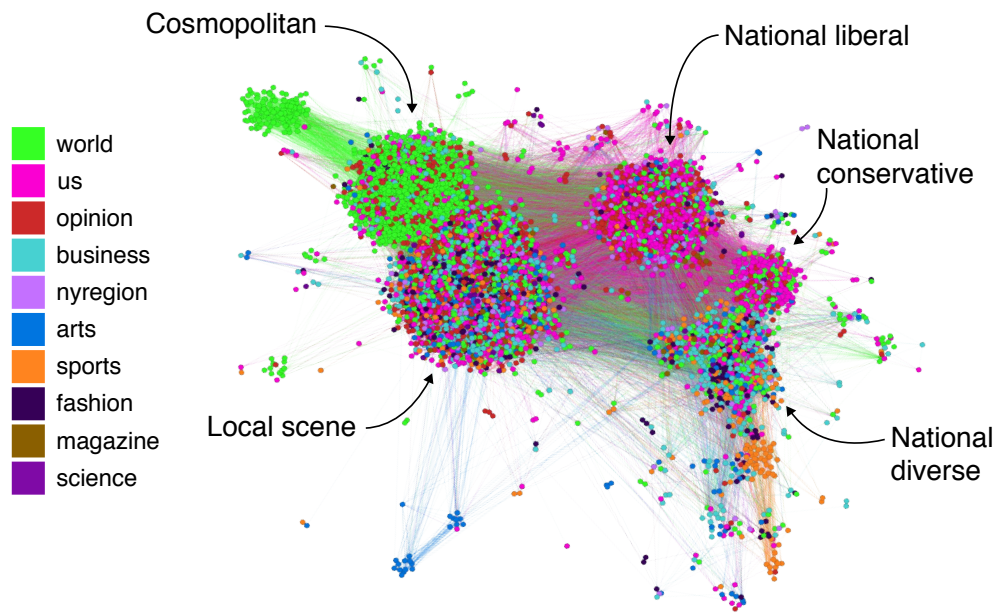}
\caption{News sharing network. Nodes are individuals who predominantly share news stories on topics given by the legend. Links are ``follow'' relationships between individuals. Cosmopolitan, local scene, national liberal, national conservative, and national diverse are tightly connected groups.
}
\label{fig:groups}
\end{figure}

Who interacts with whom affects the characteristic behavior of social systems. The rapid development of electronic social media \cite{henrikson} 
may radically change both the structure of social networks and the actions they take \cite{thompson,briggs}. 
Here we map the structure of news sharing social groups (Figure \ref{fig:groups}), revealing that they are organized around local, national and cosmopolitan concerns. The national interest group is partitioned into three subgroups: two that are politically liberal and conservative, and a third that is interested in a variety of topics, including business, sports and arts. 

Social links determine what kind of information people are exposed to, and thus how they respond to ongoing events, especially when time is a critical factor in the response. What information is selected for transmission also influences the development of perspectives. Strongly linked groups that rapidly share information may react together and with shared views \cite{mcpherson2001,sunstein2006,gilbert2009}.  The wider social propagation and influence of information is then limited by the strength of ``weak ties'' between these groups~\cite{Granovetter_1973}. 
An understanding of society not only depends on characterizing the behavior of individuals and their immediate links, but also on identifying emergent social collectives. For example, individuals might group by political orientation, economic class, race, family, profession, age, religion or geography, or by domain of interest such as business, sports, arts and politics. 

One of the universal properties of complex systems is substructure \cite{simon,DCS}, the characteristics of which are essential to a particular system's distinct qualities \cite{DCS}. 
The role of substructure is apparent in an analysis of the information flows in the human brain, where sensory, motor and other functional separations of information are key to the overall information processing. Similarly, we can expect that the differentiation across social groups will play an important role in the behaviors and function of society. 

Communication media, both peer-to-peer and broadcast, connect individuals separated in space and time. The patterns of communication change not only the experience of an individual but also the structure and behavior of our social systems. 

Before electronic communication, strongly linked social groups were primarily geographic \cite{thompson,briggs,poe,winston}.
Individual and group identities were defined by the frequent interactions of each individual with other members of the group. The variations among groups, their distinct cultures, arose from variations in local conditions and historical events. Individuals might change their associations by moving, and groups might change their location by migrating. As a way of understanding the changing nature of  human social organization, we might calibrate it to the growing frequency of longer range interactions, and the extent to which communication is one-to-one (peer-to-peer) or one-to-many, the latter being a characteristic of mass media. From messengers, to letters, telegrams, telephone, e-mail and e-messaging, peer-to-peer communications became easier and more frequent. From public presentations, to printing presses, radio, cinema, and television, mass media expanded the scope of recipients and restricted the sources of communication. The internet has merged peer-to-peer and broadcast media, increasing the flexibility of both in the form of social media. 

As the ease of long range communication increases, the dominant role of geographical association may diminish, and the process of group formation becomes motivated by acts of selection---selection by individuals of those with whom they choose to communicate. Social associations can be based upon any criteria. Released from the constraints of geography, people may choose to associate with others based upon what they consider most important and their self-identity. Instead of a local or national identity, individuals may choose a wide variety of other forms of identity. A person focused on music could choose only to communicate with others who are focused on music. Religiously focused individuals may communicate solely with others with shared religious views. Individuals who primarily value family ties may focus on those relationships. Moreover, the choices of association reinforce the form of identity that led to those associations. 

Individual choices of who to interact with result in the overall communication structure of society. 
The communication structure is then responsible for the collective behaviors that arise when news about events, and opinion formation, result in social movements and collective decision making \cite{twitterrising,crowdsourcing,wisdomcrowds}.  Understanding this structure and dynamics can be fundamental to our ability to characterize society. The availability of detailed information about social media provides a new way to advance our understanding of the dynamics of individual choices of association and social organization. Our objective is to advance this understanding by starting to characterize the spontaneous development of collective structures in social media. We focus specifically on those structures associated with news information sharing, which are the most directly relevant to societal response to news events.

Hundreds of millions of people around the world use social media and micro-blogging platforms as a real-time information source~\cite{Wu2011Who, Mendoza2010Twitter, Yu_etal:2011}. Here we identify news-sharing communities in the micro-blogging platform Twitter. Twitter users post very short (at most 140-character) posts called tweets. More than a hundred million tweets are posted every day~\cite{twitter2011}. These tweets are typically about what users do, think, or experience and want to share with other people~\cite{Java_etal:2007}. Sharing URLs and reported news constitutes a significant fraction of user activities in Twitter~\cite{naaman2010really,suh2010want}. The brevity of tweets and their rapid distribution facilitates sharing and receiving the news as events unfold \cite{Lenhart2009Twitter}. 

Twitter enables individuals general access to tweets, but also allows people to construct social connections that form primary channels of communication. Twitter users identify other users  to ``follow'' in order to be notified of their tweets. Each user can be thought of as a node in a network, and the relationships as links between them. 

A ``follow relation" in Twitter may reflect various kinds of relationships. Preexisting friends may follow each other, but there are also ways to discover others to follow within the Twitter system. Since users share what is important to them, the way the Twitter society organizes itself both reflects and influences social communication. Research has confirmed that Twitter users with similar interests tend to connect to each other~\cite{Java_etal:2007}. Analyses suggest that Twitter serves more for information sharing than for casual conversation \cite{Kwak10www, Wu2011Who}. There is also research that has associated specific Twitter users or user categories (celebrities, media, organizations, and blog writers) with specific intentions such as information providing, information seeking and friendship/conversation~\cite{Java_etal:2007,Wu2011Who}. For example, celebrities show an elevated interest in sport-related news, media accounts are more interested in US-related news, and organizations show more interest in science, technology and world news~\cite{Wu2011Who}. 

The ability to track the social relationships and the posted content in social media enables studies of the communities that are forming and their emerging collective behaviors. Here we study the overall structure of certain communities that have spontaneously formed in Twitter and we characterize their social, political, and geographic natures. The topology of a network has an important effect on dynamics of the network~\cite{newman2003structure} and its response to external perturbations \cite{baryamepstein2004}, as can be seen in epidemics \cite{colizza2006role} and information spread in social networks~\cite{lerman2010information}. However, our focus is on the characteristics of the spontaneous social groups that are formed and the insights that this gives about user self-identification, social priorities and collective social response.
 
We identify social groups formed out of users who tweet or retweet New York Times online articles. The New York Times is a major traditional news media outlet that is read by millions of people all over the world. We use three dimensions to characterize users: topics of interest, geographical locations, and self-description. The topics of interest are based on the articles that the users share. The topic category is identifiable in the article's URL, i.e. \emph{www.nytimes.com/2011/10/05/science/...} is a science related article. Geographical location is identified by analyzing the location field of user profiles in Twitter. Self-description is based on analyzing the terms the users use to describe themselves in their Twitter profile ``bio'' field. The multiple dimensions of identity provide insights into the nature of the groups that have spontaneously formed in the Twitter network. Despite its New York focus, the global reach of the New York Times enables us to consider the implications of the results for developing national and global communities. Details of the analysis are given in the appendices.

We collected tweets containing a New York Times article link over a fifteen day period, September 14 - 29, 2011. This resulted in a collection of 521,733 tweets posted by 223,950 unique users.  We considered only users who posted at least three URLs from the same category. We also obtained the ``following'' relationships (who follows whom) among these users, and constructed a graph of the users and their follow relations. The largest connected component of the graph contains  8,106 nodes and 163,850 links. Each user was labeled with the topic that s/he posted the most, with ties broken in favor of less popular topics. Details are provided in Appendix A. 

\begin{figure}[tb]
\includegraphics[width=0.8\textwidth]{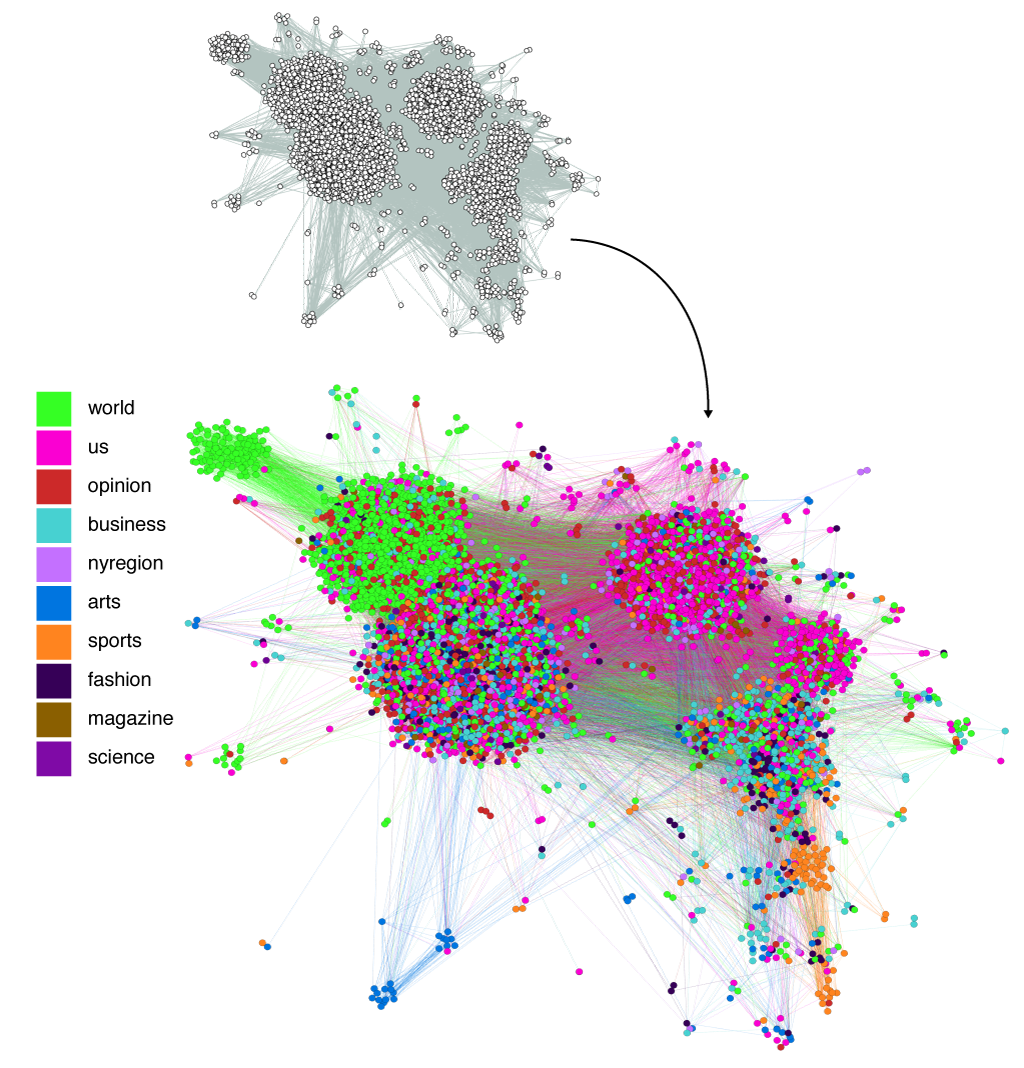}
\caption{ (top) Network of Twitter users who share New York Times online articles. Links are ``follow'' relationships between the users. The network layout is obtained by pulling followers and followees close to each other, while pushing apart unconnected nodes. The long range links that are visible occur where relatively few links connect groups of nodes that are otherwise pushed far apart. (main image) The same network layout with user nodes colored according to the topic of articles they share most. Links are shaded with the color of the follower.
}
\label{fig:network}
\end{figure}

The news sharer network we obtained is shown in Figure~\ref{fig:network}. The layout, shown at the top, is completely determined by the connections between users. Users who are connected to each other by a follow relationship are pulled together by virtual springs and those who are not are pushed apart ~\cite{Martin_etal:2011}. The visual layout then naturally identifies those nodes that belong to highly connected groups. No information about the identity or interests of the users is used to create the layout. The  clustering of the network manifests a strong social separation that can also be shown using a clustering algorithm that identifies the left and right parts of the network and further divides each to upper and lower clusters. 
Details of the layout generation and cluster generation are provided in Appendix B. 

In the main image in Figure~\ref{fig:network}, the user nodes are colored according to the article topics that users most often share in tweets. Topic colors are not randomly scattered. In some but not all cases the same colored users are clustered together. The cases where they do are consistent with the idea that users with the same interests tend to follow each other. For instance, users who are interested in the topic ``world" are clustered in the upper-left part of the network, and users who are interested in the topic ``US" are clustered in the upper-right part.

\begin{figure*}[ht]
\includegraphics[width=0.8\textwidth]{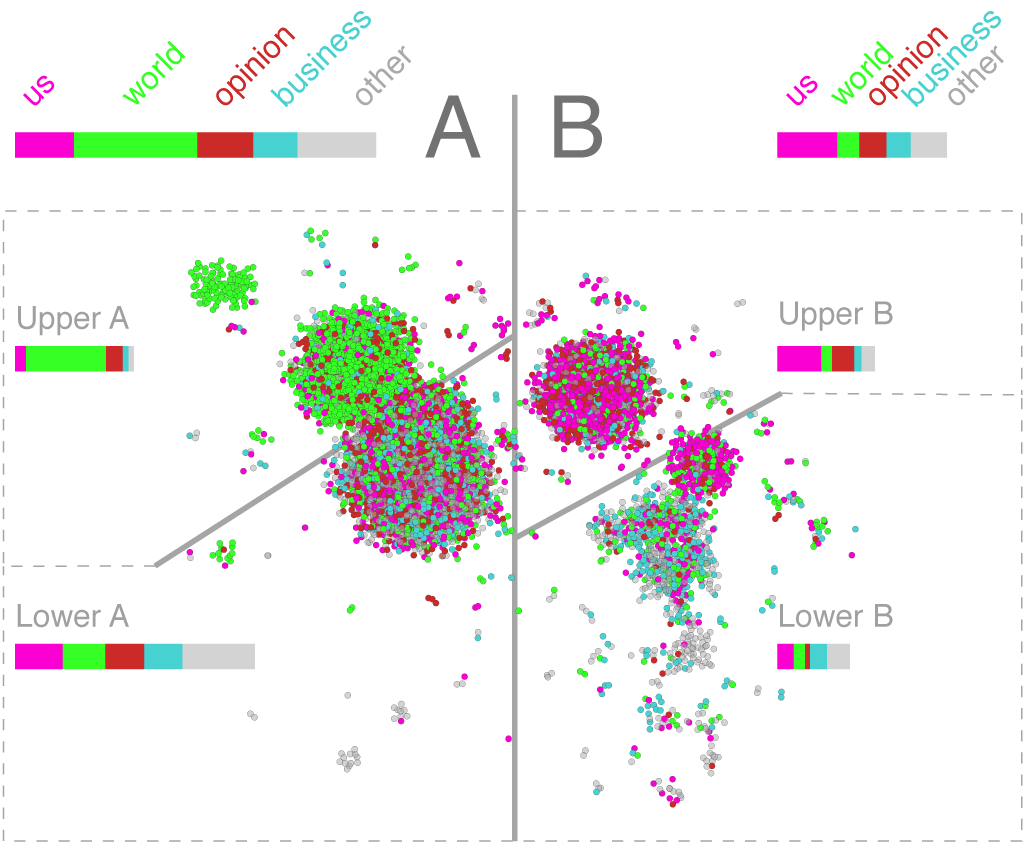}
\caption{The same network structure given in Figure~\ref{fig:network}, links are omitted for clarity. For each of the sub-communities (A, B, Upper A, Lower A, Upper B, and Lower B) the top three URL topic distributions are identified and these categories are given in the bar charts. The width of each bar-chart component is proportional to the number of users in the corresponding category. The total width of each bar chart is proportional to the total number of users in each community.}
\label{fig:clustering}
\end{figure*}

The most prevalent topics of interest for the primary clusters are shown in Figure~\ref{fig:clustering}. The division of left (A) and right (B) clusters most prominently separates those who are interested in world news (cluster A) from those interested in US-based news (cluster B). We further divide cluster A into upper and lower sub-clusters and find that the dominant topic in upper A is \category{world} news whereas lower A has a more diverse set of topics with \category{business}, \category{US}, \category{arts}, \category{fashion}, and \category{sports} being well represented in this sub-cluster.

The results show that the clusters are quite distinct and begin to reveal their nature. For example, it is not the case that each news topic has its own cluster. Instead, some topics differentiate the clusters, while others are grouped together. The clear separation between US and world news suggests that geography might be the key to the structure of the social network (A geographical location analysis is given in Appendix C). However, the sense in which geography plays a role is through the scope of interest: local, national and global.

\begin{figure*}[ht]
\includegraphics[width=1\textwidth]{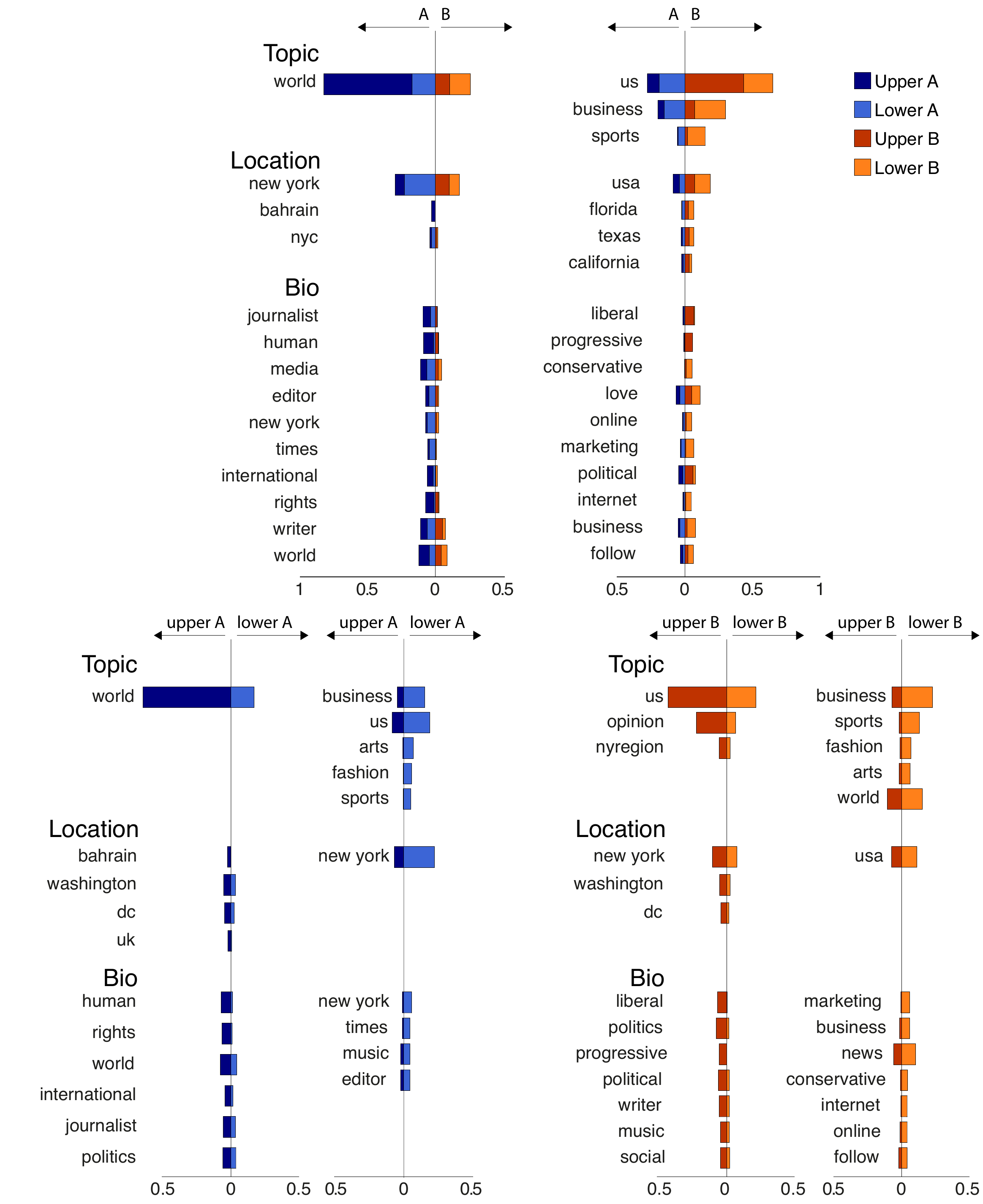}
\caption{Comparison between clusters based on topic, location and biography attributes. Each vertical axis shows the most dominant terms associated with that cluster. The top figure compares Cluster A (left) and Cluster B (right), as well as the upper and lower domains shown by distinct colors. The bottom left two axes and right two axes compare upper A to lower A, and upper B to lower B, respectively.}
\label{fig:comp_all}
\end{figure*}

Identifying the most distinctive aspects of each cluster in all three dimensions (topic, location, and biography) provides a more complete picture as shown in Figure~\ref{fig:comp_all} (Details are given in Appendix D). 

\begin{itemize}

\item Upper A is formed of users interested in international topics, who live in various cities around the world, including New York and Washington DC, are focused on human rights and politics, and may themselves be journalists. We might call them the cosmopolitan group. The small separate cluster in the upper left consists of users giving their location as Bahrain, which appear due to the Arab Spring events during the period of time we studied. 

\item Lower A is a New York oriented group, many of whom say they are located in New York, interested in a diverse set of topics including world, US, business, arts, fashion and sports. We can call them the local or ``New York scene."

\item Upper B is interested primarily in US politics, is US based, with focal locations in New York and Washington DC, and is predominantly liberal and progressive. 

\item Lower B, like Lower A, has a diverse set of topics with a stronger focus on business, but also interested in world news, sports, fashion and the arts. It is geographically spread across the US and has an apparently conservative bent. However, we found that this orientation is specific to one subgroup of Lower B.   
Figure~\ref{fig:network} shows that the Lower B group itself separates into a top right closely coupled cluster and other less strongly clumped ones. The top right group is primarily interested in US politics and self-identifies as conservative, and the rest of Lower B has a broad array of interests and is not politically distinct. 

\end{itemize}

Given the separation of the Lower B group into the conservative cluster and the rest, we have five primary clusters that can be readily identified. A simple characterization of these groups is that three are identified with New York, US and Global interests. The other two clusters are also US based but are specifically liberal and conservative in their political orientation.

Aside from the main clusters, there are small topically focused clusters of users particularly in the lower part of Figure 1, that can be seen to be primarily interested in arts or sports. 

The results reveal intriguing insights about how the network is organized. Traditional geographic groups would be associated with individual cities or countries. Freedom from geography does not eliminate entirely geographical association but results in subject driven communities that are interested in local (NY), national (US) and global (cosmopolitan) issues.
The national group subdivides into liberal and conservative political groups and a diverse but mostly business oriented group with 
sports, arts and other splinters. A person who is cosmopolitan associates with others who are cosmopolitan, and a US liberal / conservative associates with others who are US liberal / conservative, creating separated social groups with those identities. Thus, while local associations are sometimes primary, individuals may choose to associate with national or cosmopolitan groups.  A significant fraction of the population have become so strongly identified with ideological camps that those identities drive their social associations.  For those who are concerned about the polarization of society into liberal and conservative camps \cite{bishop2008,Levendusky2009}, the results have both positive and negative connotations. The groups that stratify into local, national and cosmopolitan are not politically polarized, but there are specific subgroups that are polarized into opposing camps. 

Our analysis is limited to New York Times news sharing and a specific window of time. Individual Twitter users may represent individuals or institutions. Also, the existence of specific keyword dominances is a prevalence measure and does not imply that all members or even a majority of the members of a group have that property. Nevertheless, our results manifest the strong clustering of social systems that results in specific group identity. 

Recent research has focused on user types and network topologies in social media~\cite{Nardi:2004:WWB, Java_etal:2007, Zhao_Rosson:2009, Mendoza2010Twitter, Cha_etal:2010,  Michelson_Macskassy:2010,Wu2011Who, Macskassy_Michelson:2011}. Studies have considered the distance distribution of cell phone social networks \cite{onnela2011geographic} and of a blog community \cite{liben2005geographic}. The latter manifests both short range and long range connections, which our research suggests may be due to distinct sub-groups. Our analysis also has implications for understanding the ability of individuals in social networks to intentionally transmit messages globally \cite{travers69, onnela2011geographic, watts2002}. This can be readily understood if groups that are national and cosmopolitan in orientation exist and are even weakly known by individuals of local groups. The widespread geography of members of cosmopolitan groups enable them to broadly transmit messages.

In summary, social media, and in particular Twitter, have created new opportunities for communication that are closely connected to daily lives of people, their demographics, and social and political interests. The changing communication channels also promise significant changes in the dynamics of social response to events. Social media contribute to the ability of groups to engage in activities, even including revolutions, that are underway around the world \cite{socialmediaandrevolutions}. The availability of data about these social networks and communications also provides new opportunities for understanding how social systems form and function. We used a combination of text and network analysis to study how people share information, and which patterns emerge from the social interactions and interests. The analysis reveals the natural clustering of society. We find individuals separate into groups primarily by interest in local, national or global (cosmopolitan) issues. The national group subdivides into liberal, conservative and a diverse but mostly business oriented group with sports, arts and other splinters.

We thank Karla Bertrand for helpful comments on the manuscript. This work was supported in part by ONR under grant \#N000140910516.

\section*{Appendix A: Methodology and Data Collection}

We collected tweets that contain a URL from the \href{http://www.nytimes.com}{``nytimes.com''} domain, during September 14 - 29, 2011 using the Twitter Application Programming Interface (API). This resulted in 521,733 tweets posted by 223,950 unique users. 

Categories of New York Times online articles are represented in their URLs. For example, the URL \href{http://www.nytimes.com/2011/10/05/science/space/05nobel.html}{http://www.nytimes.com/2011/10/05/science/space/05nobel.html} points to an article in the category of science. URL compression is commonly used and is accounted for in the analysis. As of September 2011, New York Times had more than 50 main categories including \category{world}, \category{us}, \category{politics}, \category{magazine} and so on. We only focused on the 10 most popular main categories in our sample (Table~\ref{tbl:categories}).

\begin{table*}[ht]
\small
\begin{tabular}{p{0.15\textwidth} p{0.25\textwidth}}
\toprule
Category & Number of URL links \\
\midrule
us & 56,252 \\
world &   52,715 \\
opinion &   43,171 \\
business &   37,582 \\
nyregion &  19,070 \\
sports &  17,309 \\
arts &  15,952 \\
magazine &  13,208 \\
science &  12,293 \\
fashion &   11,989 \\
\bottomrule
\end{tabular}
\caption{The New York Times online article categories and the number of URLs belonging to these categories, in the full sample.}
\label{tbl:categories}
\end{table*}

We limited further analysis to users who posted at least three URLs from the same category. We gathered the ``following'' relationships (who follows whom) among these users via the Twitter API, and constructed a graph representing the users and the follow relations between them. The giant component of the resulting graph contains  8,106 nodes and 163,850 links. Each user was labeled with the category that s/he posted the most. Ties were set to the less popular categories according to the number of posts in Table~\ref{tbl:categories}. 

\section*{Appendix B: Layout generation and clustering}

The spatial layout of the network given in Figure~\ref{fig:network} was determined solely by the topology of the network. We used a force-directed layout algorithm,
which heuristically optimizes the layout so that nearby unconnected nodes push each other apart and edges pull the connected nodes closer ~\cite{Martin_etal:2011}. We employed the implementation provided in the network visualization toolkit Gephi, with edge cutting parameter set to 0.65 and other parameters at their default values~\cite{ICWSM09154}.

We applied the k-means unsupervised clustering algorithm on the two-dimensional coordinates of the nodes to divide users into two communities. For given $k$, k-means heuristically identifies $k$ cluster centers such that total within cluster variance is minimal~\cite{hastie2011elements}. Using $k=2$ we obtained two clusters, A and B, which consisted of 5507 and 2599 users, respectively. To further refine the clusters, we repeated the separation of each cluster into two sub-clusters using the same process. 

\section*{Appendix C: Analysis of geographical locations}

The public profile of a Twitter user contains an optional location attribute as an unstructured text field. We analyzed the text in this field to characterize where users are from. References to the same location vary among users. For example, all of the following phrases are references to New York City: ``New York City,'' ``NYC,'' ``the big apple,'' ``Brooklyn, NY,'' ``NY City,'' ``northern manhattan.'' To overcome this problem we manually assigned each of the users into three mutually exclusive categories: New York, US (not including New York), or international (not including the US). Ambiguous cases were labeled as ``other'' with the exception that location fields which contain only a reference to the United States were labeled as US. For example, ``United States of America'' was labeled as the US, ``Northern Europe'' was labeled as international, but a single field containing the text ``California, New York, Paris'' was labeled as ``other.'' 

\begin{figure*}[ht]
\includegraphics[width=0.8\textwidth]{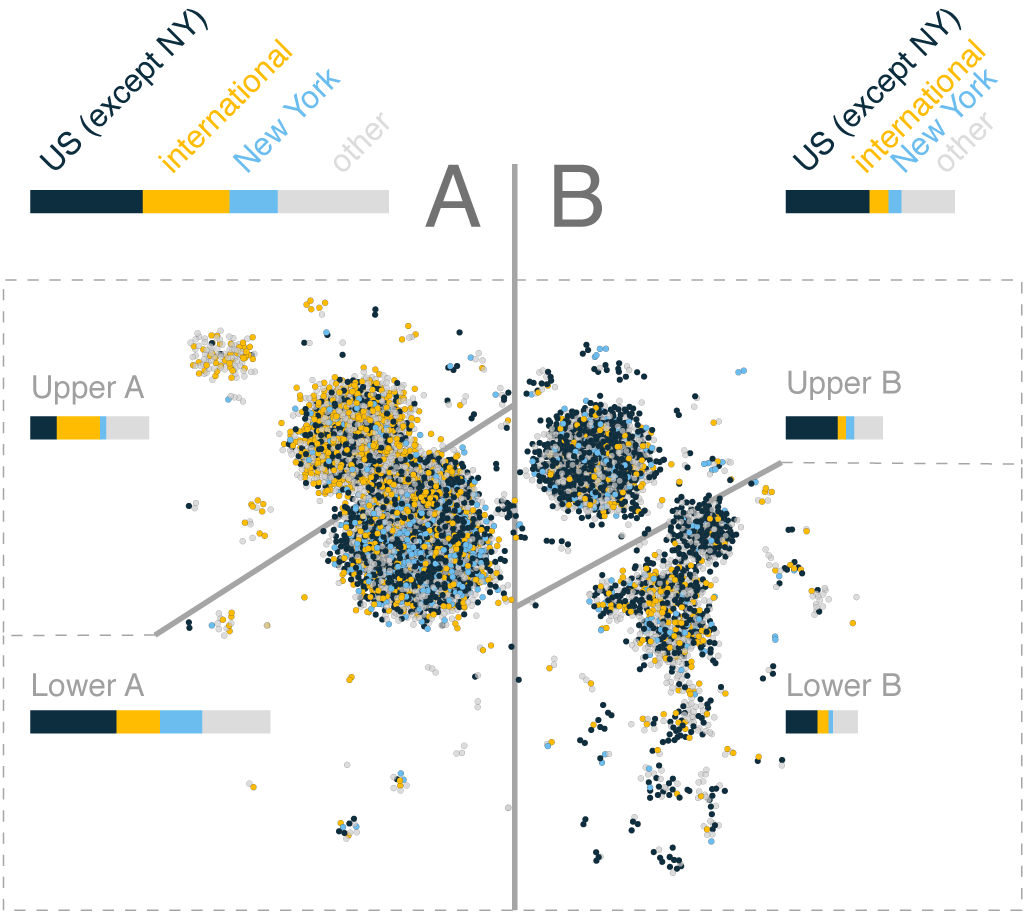}
\caption{The network structure layout given in Figure~\ref{fig:network} with edges hidden and users colored according to their geographical locations. For each cluster, the distribution of the locations are given in bar charts. The width of each bar-chart component is proportional to the number of users in the corresponding category. The total width of each bar chart is proportional to the total number of users in each cluster.}
\label{fig:us-ny-intl}
\end{figure*}

In Figure~\ref{fig:us-ny-intl}, the location distribution of each cluster is given. 

\section*{Appendix D: Analysis of overrepresented words}

The public profile of a Twitter user contains a short biography (bio) of the user as an optional, free-text field. We analyzed the words occurring in this field to characterize how they describe themselves.
Even though these fields may contain poorly characterized terms at the individual level (such as non-standard language or typos), it is possible to capture the main characteristics of communities via aggregating text-based statistics over large number of people.

We employed a bag-of-words approach, a standard text representation technique which discards the word order and represents a chunk of text as a set of words occurring in the text \cite{kao2007natural}. We converted upper to lower-case, used the space character to split the text into words and did not carry out any lemmatization -- the surface forms of the words such as ``works'' and ``working'' were counted separately. We removed any word which contained a character outside the English alphabet and two accented characters \'{e}, \`{e}, and -. We also discarded commonly used words (called stop words) such as ``and,'' ``the'' and ``of.''

We define the prevalence of a location term $w$ in a cluster $c$ as the ratio of users in $c$ who mention $w$ in their location fields, and denote it by $p_{location}(w,c)$. We denote the analogous measure for the terms mentioned in the bio field by $p_{bio}(w,c)$, and for the article topics by $p_{topic}(w,c)$. For simplicity, when we talk about all three measures, we refer to them simply as $p_{*}$. 

In Table~\ref{tbl:top-description-words}, we list the top-five words with the highest $p_{bio}$ values for the four sub-clusters. The lists give some idea about how the users in each sub-cluster identify themselves. However, the rankings of words according to the $p$ measures are dominated by the raw frequency and it is hard to compare different groups of people based on this raw frequency values. For instance some frequently used words such as ``com'' and ``news'' are found in more than one list. To compare clusters, we use the difference in prevalences of the terms in the two clusters. Given two clusters $c_1$ and $c_2$, the bias of a word $w$,  is $b_{*}(w, c_1, c_2) = p_{*}(w, c_1) - p_{*}(w, c_2)$. The sign of this measure is positive for $c_1$ cluster dominant words.

\begin{table}[ht]
\small
\begin{tabular}{p{0.15\textwidth}p{0.15\textwidth} p{0.15\textwidth}p{0.15\textwidth}p{0.15\textwidth}}
\toprule
Rank & Upper A & Lower A & Upper B & Lower B \\
\midrule
1 & news & news & politics & news \\
2 & world &   com & liberal & love \\
3 & human &   media & political & business \\
4 & rights &   writer & news & marketing \\
5 & politics &  new-york & progressive & com \\
\bottomrule
\end{tabular}
\caption{Most prevalent words in the description fields of users for each of four sub-clusters.}
\label{tbl:top-description-words}
\end{table}

To generate Figure \ref{fig:comp_all}, for each attribute (topic, location, and biography), we identified the 10 most prevalent terms in each of the four sub-clusters, and computed their bias. The most A-skewed terms are given on the left and the most B-skewed terms are given on the right, in decreasing bias order. The bar widths are proportional to the prevalence values of the terms. For instance, the term ``liberal'' is among the most dominant biography terms in cluster B; therefore it is given on the right side of Figure 1. Moreover we also see that the prevalence of ``liberal'' in Upper B is much higher than its prevalence in lower B by comparing the width of the dark orange bar to the width of the light orange bar (which is almost non-existent).

The bottom left and right parts of Figure \ref{fig:comp_all} are generated similarly, considering only lower and upper A (B) respectively.


\begin{thebibliography}{10}

\bibitem{henrikson}
J.~U. Henrikson, The growth of social media: An infographic, {\it Search Engine
  Journal\/} (8/30/2011
  \url{http://www.searchenginejournal.com/the-growth-of-social-media-an-infogr%
aphic/32788/}).

\bibitem{thompson}
J.~B. Thompson, {\it The media and modernity: A social theory of the media\/}
  (Stanford University Press, Stanford, CA, 1995).

\bibitem{briggs}
A.~Briggs, P.~Burke, {\it Social History of the Media: From Gutenberg to the
  Internet\/} (Polity Press, Cambridge, UK, 2002).

\bibitem{mcpherson2001}
M.~McPherson, L.~Smith-Lovin, J.~M. Cook, {B}irds of a feather: {H}omophily in
  social networks, {\it Annual Review of Sociology\/}, {\bf 27}, 415-444
  (2001).

\bibitem{sunstein2006}
C.~R. Sunstein, {\it Infotopia: How Many Minds Produce Knowledge\/}.

\bibitem{gilbert2009}
E.~Gilbert, T.~Bergstrom, K.~Karahalios, {B}logs are echo chambers: {B}logs are
  echo chambers, Proceedings of Hawaii International Conference on System
  Sciences 2009 (2009).

\bibitem{Granovetter_1973}
M.~S. Granovetter, {T}he {s}trength of {w}eak {t}ies, {\it {A}merican {J}ournal
  of {S}ociology\/}, {\bf 78}, 1360--1380 (1973).

\bibitem{simon}
H.~A. Simon, {\it The Sciences of the Artificial, 3rd ed.\/} (MIT Press,
  Cambridge, MA, 1999).

\bibitem{DCS}
Y.~Bar-Yam, {\it Dynamics of Complex Systems\/} (Westview Press, Boulder, CO,
  1997).

\bibitem{poe}
M.~T. Poe, {\it A History of Communications: Media and Society from the
  Evolution of Speech to the Internet\/} (Cambridge University Press, New York,
  NY, 2011).

\bibitem{winston}
B.~Winston, {\it Media Technology and Society: A History from the Telegraph to
  the Internet\/} (Routledge, New York, 2000).

\bibitem{twitterrising}
Y.~Bar-Yam, Twitter rising: Network scientists say social media are just what
  our society needs, {\it PhysOrg\/} (March 29, 2011
  \url{http://www.physorg.com/news/2011-03-twitter-network-scientists-social-m%
edia.html}).

\bibitem{crowdsourcing}
J.~Howe, {\it Crowdsourcing: Why the Power of the Crowd is Driving the Future
  of Business\/} (Crown Business, New York, NY, 2008).

\bibitem{wisdomcrowds}
J.~Surowiecki, {\it The Wisdom of Crowds\/} (Anchor Books, New York, 2005).

\bibitem{Wu2011Who}
S.~Wu, J.~M. Hofman, W.~A. Mason, D.~J. Watts, {Who says what to whom on
  Twitter}, Proceedings of the 20th International Conference on World Wide Web
  - WWW '11 705--714 (2011).

\bibitem{Mendoza2010Twitter}
M.~Mendoza, B.~Poblete, C.~Castillo, {Twitter under crisis: Can we trust what
  we RT?}, Proceedings of the First Workshop on Social Media Analytics 71--79
  (2010).

\bibitem{Yu_etal:2011}
L.~{Yu}, S.~{Asur}, B.~A. {Huberman}, {What Trends in Chinese social media},
  The 5th SNA-KDD Workshop '11 (2011).

\bibitem{twitter2011}
Twitter, Numbers (2011). \url{http://blog.twitter.com/2011/03/numbers.html}
  Fetched October 22, 2011.

\bibitem{Java_etal:2007}
J.~Akshay, S.~Xiaodan, F.~Tim, T.~Belle, {Why we Twitter: Understanding
  microblogging usage and communities}, Procedings of the Joint 9th WEBKDD and
  1st SNA-KDD Workshop 2007 (2007).

\bibitem{naaman2010really}
M.~Naaman, J.~Boase, C.~Lai, {Is it really about me?: Message content in social
  awareness streams}, Proceedings of the 2010 ACM Conference on Computer
  Supported Cooperative Work 189--192 (2010).

\bibitem{suh2010want}
B.~Suh, L.~Hong, P.~Pirolli, E.~Chi, {Want to be retweeted? Large scale
  analytics on factors impacting retweet in Twitter network}, 2010 IEEE Second
  International Conference on Social Computing (SocialCom) 177--184 (2010).

\bibitem{Lenhart2009Twitter}
A.~Lenhart, {Twitter and status updating: Demographics, mobile access and news
  consumption}, {\it Pew Internet and American Life Project\/} (2009).
  \url{http://www.pewinternet.org/Reports/2009/Twitter-and-status-updating.asp%
x} Fetched October 22, 2011.

\bibitem{Kwak10www}
H.~Kwak, C.~Lee, H.~Park, S.~Moon, {W}hat is {T}witter, a {s}ocial {n}etwork or
  a {n}ews {m}edia?, {WWW '10: Proceedings of the 19th International Conference
  on World Wide Web} 591--600 (2010).

\bibitem{newman2003structure}
M.~Newman, The {s}tructure and {f}unction of {c}omplex {n}etworks, {\it {SIAM}
  {R}eview\/} 167--256 (2003).

\bibitem{baryamepstein2004}
Y.~Bar-Yam, I.~Epstein, Response of {c}omplex {n}etworks to {s}timuli, {\it
  {PNAS}\/}, {\bf 101}, 4341--5 (2004).

\bibitem{colizza2006role}
V.~Colizza, A.~Barrat, M.~Barth{\'e}lemy, A.~Vespignani, The {r}ole of the
  {a}irline {t}ransportation {n}etwork in the {p}rediction and {p}redictability
  of {g}lobal {e}pidemics, {\it PNAS\/}, {\bf 103}, 2015 (2006).

\bibitem{lerman2010information}
K.~Lerman, R.~Ghosh, {I}nformation contagion: {A}n empirical {s}tudy of the
  {s}pread of {n}ews on {D}igg and {T}witter {s}ocial {n}etworks, {P}roceedings
  of 4th {I}nternational {C}onference on {W}eblogs and {S}ocial {M}edia
  ({ICWSM}) (2010).

\bibitem{Martin_etal:2011}
S.~Martin, W.~M. Brown, R.~Klavans, K.~W. Boyack, Openord: An open-source
  toolbox for large graph layout, Conference on Visualization and Data
  Analysis, {\bf 7868}, 786806 (2011).

\bibitem{bishop2008}
B.~Bishop, {\it {T}he big sort: {W}hy the clustering of like-minded {A}merica
  Is tearing us apart\/} (Houghton Mifflin Harcourt, Boston, MA, 2008).

\bibitem{Levendusky2009}
M.~S. Levendusky, {\it {T}he Partisan Sort: How Liberals Became Democrats and
  Conservatives Became Republicans\/} (University of Chicago Press, Chicago,
  IL, 2009).

\bibitem{Nardi:2004:WWB}
B.~A. Nardi, D.~J. Schiano, M.~Gumbrecht, L.~Swartz, Why we blog, {\it
  Commununications of the ACM\/}, {\bf 47}, 41--46 (2004).

\bibitem{Zhao_Rosson:2009}
D.~Zhao, M.~B. Rosson, How and why people twitter: the role that micro-blogging
  plays in informal communication at work, Proceedings of the ACM 2009
  International Conference on Supporting Group Work 243--252 (2009).

\bibitem{Cha_etal:2010}
M.~Cha, H.~Haddadi, F.~Benevenuto, K.~P. Gummadi, {Measuring user influence in
  Twitter: the million follower fallacy}, {Proceedings of International AAAI
  Conference on Weblogs and Social Media} (2010).

\bibitem{Michelson_Macskassy:2010}
M.~Michelson, S.~A. Macskassy, Discovering users' topics of interest on
  {T}witter: {A} first look, Proceedings of the Fourth Workshop on Analytics
  for Noisy Unstructured Text Data 73--80 (2010).

\bibitem{Macskassy_Michelson:2011}
S.~A. Macskassy, M.~Michelson, Why do people retweet? {A}nti-homophily wins the
  day!, {Proceedings of the Fifth International AAAI Conference on Weblogs and
  Social Media} (2011).

\bibitem{onnela2011geographic}
J.~Onnela, S.~Arbesman, M.~Gonz{\'a}lez, A.~Barab{\'a}si, N.~Christakis,
  Geographic constraints on social network groups, {\it PLoS one\/}, {\bf 6},
  e16939 (2011).

\bibitem{liben2005geographic}
D.~Liben-Nowell, J.~Novak, R.~Kumar, P.~Raghavan, A.~Tomkins, Geographic
  routing in social networks, {\it Proceedings of the National Academy of
  Sciences of the United States of America\/}, {\bf 102}, 11623 (2005).

\bibitem{travers69}
J.~Travers, S.~Milgram, An experimental study of the small world problem, {\it
  Sociometry\/}, {\bf 32}, 425-443 (1969).

\bibitem{watts2002}
D.~J. Watts, P.~S. Dodds, M.~E.~J. Newman, {I}dentity and search in social
  networks, {\it Science\/}, {\bf 296}, 1302-1305 (2002).

\bibitem{socialmediaandrevolutions}
G.~Lotan, M.~Ananny, D.~Gaffney, D.~Boyd, The revolutions were tweeted:
  {I}nformation flows during the 2011 {T}unisian and {E}gyptian revolutions,
  {\it International Journal of Communication\/}, {\bf 5}, 1375--1405 (2011).

\bibitem{ICWSM09154}
M.~Bastian, S.~Heymann, M.~Jacomy, {G}ephi: {A}n open source software for
  exploring and manipulating networks, {I}nternational {AAAI} Conference on
  Weblogs and Social {M}edia (2009).

\bibitem{hastie2011elements}
T.~Hastie, R.~Tibshirani, J.~Friedman, J.~Franklin, {\it The {e}lements of
  {s}tatistical {l}earning: {D}ata {m}ining, {i}nference and {p}rediction\/}
  (Springer Verlag, 2011).

\bibitem{kao2007natural}
A.~Kao, S.~Poteet, {\it Natural language processing and text mining\/}
  (Springer Verlag, 2007).

\end{thebibliography}
\end{document}